# Wireless Sensor Network MAC Energy - efficiency Protocols: A Survey

Ghassan Samara

Computer Science Department, Zarqa University, Zarqa- Jordan

gsamarah@zu.edu.jo

Abstract— Energy Efficiency in wireless sensor networks is an important topic in which the nodes rely on battery power, and efficient energy usage is a key issue for sensitive applications that require long working times. This stimulates many scientists at all levels of communication protocols Medium Access Control (MAC) who control the use of the wireless transmitter and receiver unit to create new protocols.
Many protocols were suggested that primarily take energy efficiency as the primary objective of sustaining the function of the network for as long as possible into account with different objectives for wireless sensor networks.
This paper will look at some of these energy efficiency protocols.

## I. INTRODUCTION

The advances at the sensor level resulted in cost reduction for a sensor consisting of a built-in processor and a transceiver [1]. Low sensor node power capabilities lead to little communication coverage compared to other mobile devices, which need a large number of sensors to cover the target area successfully. In addition, they are close and ineffective to load or replace the node batteries, unlike other wireless networks [2]. This is reduced due to the demanding environment in which those nodes are used, which increases the efficiency of the primary energy objective in the design of wireless sensor networks to improve the life of sensor nodes [3].

The MAC network of wireless sensors must be able to face many major challenges due to the wireless network's popular existence. The hidden terminals can, for example, cause a collision and widespread wireless usage on one site can cause significant inconsistencies, thereby weakening the capacity of the sensor. Many other wireless interference sources too, including other wireless devices such as a malicious node, carry out an aggressive noise attack that constantly transmits to prevent other nodes from accessing the channel [4].

However, medium resolution within a dense duty-low cycle node network is a serious concern that must be resolved in an effective energy way. Finally, a protocol survey of MAC in relation to protocols and future recommendations for researchers on open topics that have not been well studied [5].

When more than one data packet occurs in a wireless sensor node at the same time, the collision packets are discarded, and retransmissions lead to waste energy [6]. They also consume when a sensor node listens to data packets from other nodes received and also when a node sends a data packet to another node not ready to receive data, all of which leads to a waste of energy. Hence, The design of MAC protocols solves these problems to preserve a key feature of the energy wireless sensor network [7].

This paper discusses a series of MAC protocols designed to solve data transfer problems and to improve energy efficiency and the contrast between them.

## II. LITERATURE REVIEW

The problems in the data transmission process and the resulting energy waste in wireless sensor networks would require a solution to these problems and solutions found on a group of MAC protocols like [S-MAC] Protocol [8], This works by directing node to idle during transmissions to other nodes to minimize energy consumption [9]. The authors work on variations between overload, time and location for data transmission that reduces power waste in the T-MAC Protocol [10, 11]. The protocol (TEEM) [12] is also based on the principle of listening during broadcasting and on the transition to unfailing times. Who inspires the (S-MAC Protocol).

A protocol (EM-MAC) has been established[13] which encourages the use of the channel and the efficiency of the transfer of information to some part of the channels. The protocol (LMAC)[14], developed explicitly for wireless sensor networks, has been developed. (TDMA)[15] The protocol used for incident-free communication and the key purpose of this protocol to resolve a physical layer network overall issue.

## III. MAIN DESIGN OBJECTIVE IN WSN

Many sensor nodes are for particular applications, and there is a cluster of specifications for each application. In the design of wireless sensor networks, which is a wholly or partly necessary aspect of network design, we will show the key objectives [16, 17] see figure 1:

Small node size: These nodes are typically distributed in large numbers over the harsh environment or military areas, which greatly promote the distributed process and minimize energy consumption and sensor node costs.

Low node cost: Therefore, reducing costs for wireless sensor nodes is essential in the process of spreading nodes in a hard environment, as they minimize the network's overall cost.

Low power consumption: Due to its design, recharging and repairing processes is very difficult and could be impossible. To do this, minimizing electrical energy consumption when transmitting data helps to extend network life for a long time. Wireless sensors may get their power from the internal battery.

Reliability: When receiving the data and interference sensor networks, error corrections must be controlled through wireless controls, and the data processes must be held proper to ensure trust in the presence of transmitter channel problems.

Scalability: Wireless sensor network can handle different network dimensions, depending on their applications and can rely on hundreds to thousands of sensor nodes.

Self-reconfigurability: The nodes of sensors shall be able to organize themselves independently and also be able to reorganize any network changes, for example, when one or more nodes are dead.

Channel utilization: The nodes of the sensors are reserved for bandwidth. Most communication protocols are intended to improve the effective use of the channel.

Adaptability: Failure of the nodes to connect a cluster or to transfer to another cluster results in a shift in the architecture of the network and the network should then be adapted to these design changes.

Security: A mechanism to protect such data from theft or to modify the original material using protocols should be in place to retain the transmitted data, and that can form part of critical applications.

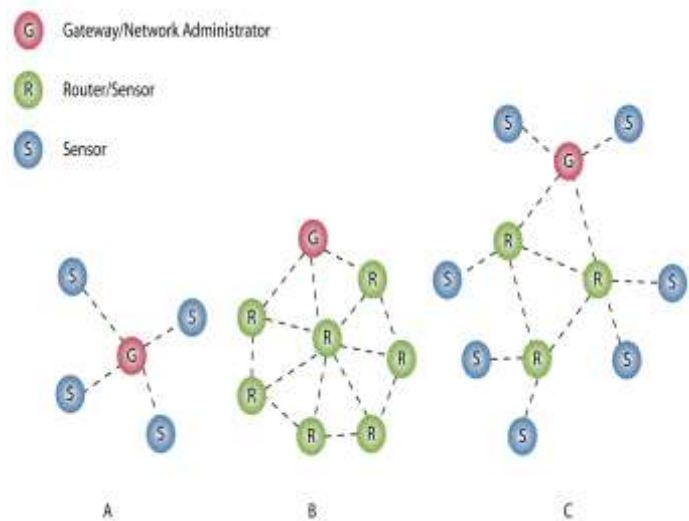

Fig.1. WSN design

Adequate MAC protocol for Wireless Sensor Networks must be built with the following attributes [18].

Firstly, energy efficiency protocols have to be identified, which are essential for extending network life. Scalability and adaptability to changes are also important features.

Changes for effective adaptation should be dealt with quickly and effectively, changes in network size, node density and topology, for instance. Changes in network property occur due to various factors, some of which are limited node life, the addition of new nodes to the network and changes of interference can alter network connectivity. In sensor networks, other essential attributes, including latency, passage and the usage of bandwidth, may be minor.

Such network adjustments should be modified in a successful MAC protocol. Unlike other wireless networks, fairness between the sensor nodes is typically not a design objective, since all sensor nodes have a similar mission.

### IV. COMMUNICATION TYPE IN WSN

In wireless sensor networks identified by [19], there are three types of communication patterns that broadcast, converge cast and local gossip.

The first form of the pattern is broadcast. It is normally used to send any data to all network sensor nodes. The information generated by broadcasting can also include queries for sensor query processing architectures, software updates for sensor-based nodes, and control packets for the entire system. It must also be used by a primary station (sink) [20, 21].

The communication pattern of broadcast type packet should not be adversely affected by the broadcast type. Both network nodes are intended recipients for the initial period, while the last nodes for the transmitter are called recipients for the transmitter node's contact range [22].

Local gossip is considered the second form of the communication pattern. In some stories, sensors that detect a medium interact locally with each other, when a sensor sends a message within the same region to the surrounding nodes. The third pattern of communication called converge casting, wherever a group of sensors interacts with a single sensor, is to send what the sensors see about the medium which has been identified to the information center. The target node could be the head of a cluster—either a center or base station for data fusion [23, 24].

### V. MAC PROTOCOLS

Sensor-MAC (S-MAC), the key concept in this protocol is When neighbouring nodes form virtual clusters to create a mutual sleep plan. If two adjacent nodes stay in two separate virtual clusters, they wake up in both clusters' listening hours. The possibility to obey two separate schedules is a challenge to the S-MAC algorithm. Passive listening and overhearing contribute to energy usage above the average level. Periodic SYNC packets should be transmitted to the immediate neighbours to arrange exchanges. The synchronization time is called each node to send a SYNC packet, see figure 2.

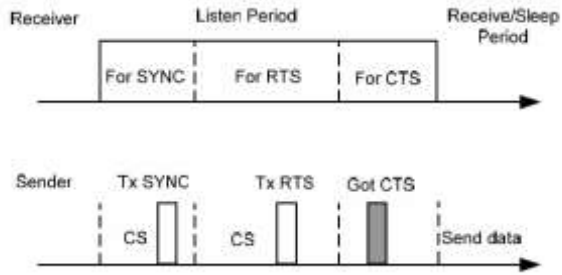

Fig.2. S-MAC-Messaging-Scenario

The sender-receiver communication model is explained in Figure 1. The carrier's definition as CS in the figure usually ensures the avoidance of collision. In addition, RTS / CTS packet exchanges are used in Unicast data packets. In this technique, the overall contact at the cost of medium access unfairness could be reduced by saving resources, as S-MC has an essential feature: the concept of transferring messages when long messages are separated and sent to a burst. Since every node has its sleep scheduling, periodic sleep, in particular with multi-hop algorithms, can result in high latency. The word sleep delay shown in [2] describes the lag resulting from normal sleep. The adaptive listening approach is recommended for enhancing sleep time and thus, the overall latency. The node that overhears the transmissions of the neighbor wakes up at the end of the transmission for a short time. Therefore, if the neighboring node may relay information instantly, the node is the next-hop node. The last stage is known for the RTS / CTS packet period region.

Timeout-MAC (T-MAC) as indicated earlier, Static S-MAC sleep listen cycles lead to high latency and lower efficiency. The best proposal to boost the weak S-MAC protocol results during changing traffic loads is the time-out MAC (T-MAC) [3]. In the T-MAC, TA is listening to the end of the duration where no action has taken place for a time span.

Change of load in sensor networks is anticipated as more traffic is needed in the nodes which are closer to the sink.

There are two triggers of the issue of early sleekness: variable loads, and when the listening intervals in virtual clusters are synchronized, see figure 3.

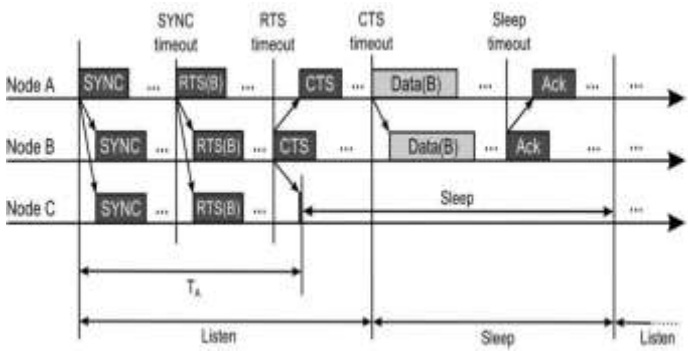

Fig. 3. Timeout-MAC (T-MAC)

Traffic Aware, Energy Efficient MAC protocol for Wireless Sensor Networks (TEEM) The TEEM protocol is an S-MAC protocol extension. The hearing interval is set, while the listening interval depends on the traffic in the TEEM protocol. All nodes will disable their radio much earlier in the TEEM protocol if there is no transfer of data. In addition, a separate RTS transmission is omitted. Every hearing interval, as in the S-MAC protocol, is divided into two sections in relation to the three components. The node sends a SYNC packet when there is some data message in the first portion of the listening interval (SYNC data). If the node does not have a data request, in the second part of its hearing interval, it will send a SYNC packet (SYNC No Data). If a node does not receive SYNC data in the first part of its interval, and does not have any data to send it will not send SYNC information to the second part of the interval of its listen. If it does not receive SYNC data, the second part of its listening interval will have no data to send it. When a node receives a new SYNC RTS, it turns the radio off and falls asleep before the next interval of listening begins. In the second part of its listening interval, the intended recipient will send CTS. The efficiency assessment of TEEM protocol reveals that TEEM sleeps a higher than S-MAC and that S-MAC control packs have a higher percentage of sleeping than the TEEM control packs.

In comparison to SMAC and IEEE 802.11, energy consumption in TEEM is the lowest. The interval is only dependent upon local traffic, node traffic itself and neighbouring node because listening decreases and it does not take account of traffic across the entire network which increases the latency and reduces energy consumption by reducing the hearing interval within the TEEM.

Multichannel Energy-Efficient MAC Protocol EM-MAC Is known as the asynchronous MAC protocol multichannel duty cycling. It has several features, where node synchronization is not required, no common control channel is not used, and channel and wake up schedule are not specifically exchanged. Instead, each node decides its own pseudorandom channel switching and wake-up times independently. An average sender meets the recipient by predicting the wake up channel of the recipient and wake up time based on the transmitter's knowledge of the pseudorandom role of the receiver used to establish its wake-up channels and times. By allowing a sender to reach a receiver precisely and easily, EM-MAC achieves extremely high energy efficiency. A transmitter in EM-MAC awakes on the expected receiver wake-up channel shortly before the recipient completes the transmission of the packet and quickly returns to its sleep state, minimizing passive listening and hearing. No special radio hardware is required for EM-MAC, see figure 4.

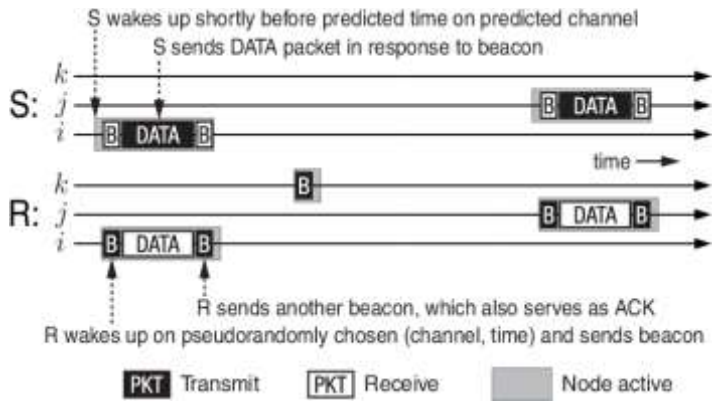

Fig. 4. Multichannel Energy-Efficient MAC Protocol EM-MAC

(Lightweight Medium Access Protocol), LMAC Built for WSN in particular. Furthermore, the LMAC protocol uses TDMA for collision-free communication. The collision-free transmission can be ensured by choosing a slot number not used within a double-hop neighbourhood, such as frequency reused in mobile networks). To do this, information broadcast in the control section provides an overview of which slots are controlled by the one-hop neighbours of the node being transmitted (i.e. the owner of the slot). New 18 Embedded Systems manual nodes that join the network can read all traffic control parts for a full structure. When OR-sets are put, they may specify, and sleeping places are still free. The new node has randomly selected a slot which needs the new node through the sending of control information. Collisions in slot selection impact Garbled control parts. A node detects a collision of this kind, transmits the entangled slot number to its control area that is overheard by unfortunate new nodes that then reverse the selection process and repeat it, see figure 5.

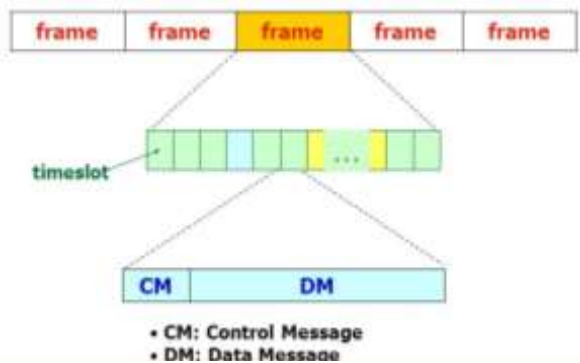

Fig. 5. Lightweight Medium Access Protocol (LMAC)

## VI. CONCLUSION

The aim of the design of MAC protocols is to extend the wireless sensor network life by safeguarding maximum sensor node life by enhancing energy consumption performance. The guideline algorithm should be designed to reduce energy consumption during data transmission.

And this survey helps to explain how these protocols operate, which may be the starting point for new research in the future.

REFERENCES:

1. Khatari, M. and Samara, G., 2015. Congestion control approach based on effective random early detection and fuzzy logic. MAGNT Research Report Vol.3 (8). PP: 180-193.
2. Samara, G., Alsalihy, W.A.A. and Ramadass, S., 2011. Increasing Network Visibility Using Coded Repetition Beacon Piggybacking. World Applied Sciences Journal, 13(1), pp.100-108.
3. Rai, Anuradha, Suman Deswal, and Parvinder Singh. 2016. "MAC Protocols in Wireless Sensor Network: A Survey." International Journal of New Innovations in Engineering and Technology 5.1: 95-101.
4. Samara, G. and Alsalihy, W.A.A., 2012. June. A new security mechanism for vehicular communication networks. International Conference on Cyber Security, Cyber Warfare and Digital Forensic (CyberSec) (pp. 18-22). IEEE
5. Samara, G., Sureswaran, R., Wafaa, A.H. and Salihy, A.I., 2010. Safety message power transmission control for vehicular ad hoc networks [J]. Computer Science, 6(10), pp.1027-1032.
6. Samara, G. and Alsalihy, W.A.H.A., 2012. Message broadcasting protocols in VANET. Information Technology Journal, 11(9), pp.1235-1242.
7. Samara, G., Al-Salihy, W.A. and Sures, R., 2010. Efficient certificate management in VANET. In 2010 2nd International Conference on Future Computer and Communication.
8. Ye, W., Heidemann, J., & Estrin, D. 2002. An energy-efficient MAC protocol for wireless sensor networks. In INFOCOM 2002. Twenty-First Annual Joint Conference of the IEEE Computer and Communications Societies. Proceedings. IEEE (Vol. 3, pp. 1567-1576). IEEE.
9. Karaoglu, B., Numanoglu, T., Tavli, B. and Heinzelman, W., 2015. Energy Efficient Real-Time Distributed Communication Architectures for Military Tactical Communication Systems. In Enabling Real-Time Mobile Cloud Computing through Emerging Technologies (pp. 35-82). IGI Global.
10. Van Dam, T., & Langendoen, K. 2003. An adaptive energy-efficient MAC protocol for wireless sensor networks. In Proceedings of the 1st International Conference on Embedded networked sensor systems (pp. 171-180). ACM.
11. Althobaiti, Ahlam Saud, and Manal Abdullah. 2015. "Medium access control protocols for wireless sensor networks classifications and cross-layering." Procedia Computer Science 65 : 4-16.
12. Suh, C., & Ko, Y. B. 2005. A traffic aware, energy-efficient MAC protocol for wireless sensor networks. In Circuits and Systems, 2005. ISCAS 2005. IEEE International Symposium on (pp. 2975-2978). IEEE.
13. Tang, L., Sun, Y., Gurewitz, O., & Johnson, D. B. 2011. EM-MAC: a dynamic multichannel energy-efficient MAC protocol for wireless sensor networks. In Proceedings of the Twelfth ACM International Symposium on Mobile Ad Hoc Networking and Computing (p. 23). ACM.
14. Van Hoesel, L. and Havinga, P., 2004, June. A lightweight medium access protocol (LMAC) for wireless sensor networks. In 1st Int. Workshop on Networked Sensing Systems (INSS 2004).
15. Pei, G., & Chien, C. 2001. Low power TDMA in large wireless sensor networks. In Military Communications Conference, 2001. MILCOM 2001. Communications for Network-Centric Operations: Creating the Information Force. IEEE (Vol. 1, pp. 347-351). IEEE.
16. Agrawal, D.P. and Manjeshwar, A., 2002, April. A hybrid protocol for efficient routing and comprehensive information retrieval in wireless sensor networks. In Proceedings of the 2nd International Workshop on Parallel and Distributed Computing Issues in Wireless Networks and Mobile computing.
17. Liu, Xuxun. 2012. "A survey on clustering routing protocols in wireless sensor networks." Sensors 12.8: 11113-11153.
18. Demirkol, Ilker, Cem Ersoy, and Fatih Alagoz. 2006. "MAC protocols for wireless sensor networks: a survey." IEEE Communications Magazine 44.4: 115-121.
19. S.S., Kulkarni, 2004. "TDMA services for Sensor Networks," Proceedings of 24th International Conference on Distributed Computing Systems Workshops, Pages:604 – 609, 23-24.

20. Samara, G. and Al-okour, M., 2020. Optimal Number of Cluster Heads in Wireless Sensors Networks Based on LEACH. International Journal of Advanced Trends in Computer Science and Engineering, ISSN 2278-3091, Volume 9, No.1.
21. Samara, G., Al Besani, G., Alauthman, M. and Al Khaldy, M., 2020. Energy-Efficiency Routing algorithms in Wireless Sensor Networks: a Survey. International Journal of Scientific & Technology Research, Vol 9, ISSUE 1.
22. Samara, G. and Blaou, K.M., 2017, May. Wireless sensor networks hierarchical protocols. In 2017 8th International Conference on Information Technology (ICIT) (pp. 998-1001). IEEE.
23. Samara, G. and Aljaidi, M., 2019. Efficient energy, cost reduction, and QoS based routing protocol for wireless sensor networks. International Journal of Electrical & Computer Engineering (2088-8708), 9(1).
24. Samara, G. and Aljaidi, M., 2018. Aware-routing protocol using best first search algorithm in wireless sensor. The International Arab Journal of Information Technology, 15(3A), pp.592-598.